\documentclass[11pt,a4paper]{article}
\usepackage{graphicx}
\usepackage{amsmath}
\usepackage{bm}

\def\itmb{\begin{itemize}}
\def\itme{\end{itemize}}
\def\enmb{\begin{enumerate}}
\def\enme{\end{enumerate}}
\def\eqnb{\begin{equation}}
\def\eqne{\end{equation}}

\title{The optimization of paths in the $R^{3,1}$ space time by \\Markov Chain Monte Carlos}

\author{Sadataka Furui$^A$ and Serge Dos Santos$^B$  \\
$^A$ Faculty of Science and Engineering, Teikyo University\\
2-17-12 Toyosatodai, Utsunomiya, 320-0003 Japan {\thanks
{\textit{E-mail address:} furui@umb.teikyo-u.ac.jp}}\\
$^B$ INSA Centre Val de Loire; Université de Tours,\\ INSERM, Imaging Brain \& Neuropsychiatry iBraiN U1253
 F-41034 Blois Cedex, France {\thanks {\textit{E-mail address:} serge.dossantos@insa-cvl.fr}}
}

\begin{document}
\maketitle
\begin{abstract}
We propose a method to obtain the optimal weight function of 9 paths in (3+1)D space-time whose length is less than or 
equal to $2\times (6+2)$ lattice units. The factor 2 comes from inclusion of opposite direction path or time reversed paths. There are $2\times 2$ time shifts, which we assume that they can be regarded as stochastic Markov processes. 
We prepare the input 9D vector ${\bf X}$ and a $9\times 9$ matrix ${\bf W}$ and a bias vector ${\bf b}$, and consider affine transformations ${\bf Z}^{(h)}={\bf X}^{(in)}{\bf W}^{(h)T}+{\bf b}^{(h)}$ and ${\bf A}^{(h)}=\sigma({\bf Z}^{(h)})$ from an input layer to a hidden layer,  the hidden layer to another hidden layer and from the hidden layer to an output layer, using the transformation
${\bf Z}^{(x)}={\bf A}^{(h)}{\bf W}^{(out) T} +{\bf b}^{(x)}$ and ${\bf A}^{(x)}=\sigma({\bf Z}^{(x)})$.
 
By choosing the matrix ${\bf W}$ a diagonal matrix, and introducing the information of action of the 9 paths, a simple Monte Carlo simulation yields actions on a 2D plane spanned by $e_1, e_2$ for a fixed $u_2=j_2 e_2$ as a function of $u_1=j_1 e_1$. The action at high momentum region has small fluctuations, but at small momentum region, has large fluctuation. Generalizing $\bf W$ including mixing of paths, we search the optimal weight function using the Machine Learning (ML) techniques. For fixed point actions, actions of the output layer are defined by the output of final hidden layer
\end{abstract}

\maketitle

\section{Introduction}
Recently, application of quaternions in engineering and physics has been intensively discussed\cite{MFLBCB23,FS01,USVDV09,UVDV10,Bridge17}.

In the Time Reversal Based Nonlinear Elastic Wave Spectroscopy (TR-NEWS), ultrasonic wave profile was expanded in 3rd order polynomials, and the wave strength of each order was mapped to quaternion bases. From convolution of a ultrasonic wave and its time reversed wave, propagating in 2D planes, anomalous scattering positions are searched\cite{GCDSBM07,DSP08,DS10,LSDS14}.   Quaternions which are elements of Clifford algebra are applied in signal and image processing\cite{MFLBCB23}. Felsberg and Sommer\cite{FS01} proposed for producing  monogenic signal from (2+1)D signal using quaternions.

Up to (2+1)D, Non Destructive Testing (NDT) using quaternion bases was successful.  We called propagation of solitonic wave on a 2D plane spanned by $e_1,e_2$ as A-type,  on a (2+1)D spanned by $e_1,e_2,e_1\wedge e_2$ as B-type.

In medical image processing, propagation of 3D materials with hysteresis effects are important. Propagation of solitonic wave in (3+1)D, which we call C-type cannot be described by quaternions, and Dirac showed a prescription of applying Lorentz transformation to quaternions\cite{Dirac45}. Although Dirac wrote schemes of bi-quaternions is not of any special interest in mathematical theory, as compared to quaternions, it has interesting physical properties. 

In Clifford algebra, the mapping $j:{\bf R}^{3,1}\to M_2({\bf H})$ proposed by Garling\cite{Garling11} is
\begin{equation}
j ({\mathcal A}^+_{3,1})=\left(\begin{array}{cc}
a_1 +a_2{\bf k}& b_1{\bf i}+b_2{\bf j}\\
c_1{\bf i}+c_2{\bf j}&d_1+d_2{\bf k}\end{array}\right),\nonumber
\end{equation}
where $a_i,b_i,c_i ,d_i\quad( i=1,2)$ are real.

In the Quantum Chromo Dynamics(QCD) lattice simulation\cite{DGHHN95}, here are 7 C-type paths. In bi-quaternion bases, there are additional 2 paths along the time direction, and we consider 9 paths.

The 6 paths $L19,L20,L21,L21',L22,L22'$ do not contain the path along the $e_1\wedge e_2$ at the beginning or the ending, while $L23,L24,L25$ contain the path along the $e_1\wedge e_2$ at the beginning or the ending, where $e_1$ and $e_2$ are the unit vector spanning a 2D plane. The path $L25$ contains the paths along $e_1,e_2$ and $e_1\wedge e_2$.

In \cite{SFDS23b, SF23c}, we considered 7 paths in (3+1)D that contain hysteresis effects. In this work, we double the length of the paths to be less than or equal to 16 lattice unit and propose a method to detecting hysteresis effect by comparison with experiments. 

The paths of 16 steps are summarized in Table 1 and Table 2. In quaternion basis $q=q_0 e_0+q_1e_1+q_2 e_2+q_3 e_3$ and $\bar q=q_0 e_0-q_1 e_1-q_2 e_2-q_3e_3$, we take the ${\bf x}=x e_1$,${\bf y}=y e_2$,${\bf z}=z e_3$, where $q_i,x,y,z \in {\bf R}$
 
Biquaternions are $e_i e_j$ $i,j\in \{0,1,2,3\}$, and when $i,j \in\{1,2,3\}$, $e_ie_j=\epsilon^{ijk}e_k$.  
$\epsilon^{123}=-\epsilon^{213}=1$, $\epsilon^{312}=-\epsilon^{132}=1$ and $\epsilon^{231}=-\epsilon^{321}=1$. 

In the Table 1 and 2, the bi-quaternion basis $e_ie_j$ are denoted as $ij$, directions of the wave front along the path are $x,y,z,t$.
Backward propagations are $-x,-y,-z,-t$. 

{\small
\begin{tabular}{r|cccccccccccccccc}
 step & 1&2&3 &4&5&6&7&8&9&10&11&12&13&14&15&16\\
\hline
L19&x&y&z&t&-z&-t&-x&-y&-x&-y&-z&-t&z&t&x&y\\
&23&31&12&24&-12&-24&-23&-31&-23&-31&-12&-24&12&24&23&31\\
\hline
L20&x&y&z&t&-z&-y&-x&-t&-x&-y&-z&-t&z&y&x&t\\
&23&31&12&24&-12&-31&-23&-24&-23&-31&-12&-24&12&31&23&24\\
\hline
L25&x&y&z&t&-x&-y&-z&-t&-x&-y&-z&-t&x&y&z&t\\
&23&31&12&24&-23&-13&-12&-24&-23&-31&-12&-24&23&13&12&24\\
\end{tabular}
}

 Table 1. Directions of the wave front of paths L19, L20,  L25 

{\small
\begin{tabular}{c|cccccccccccccccc}
step&1&2&3&4&5&6&7&8&9&10&11&12&13&14&15&16\\
\hline
L21 &x&y&z&t&-z&-x&-t&-y&-x&-y&-z&-t&z&x&t&y\\
 &23&31&12&14/24&-12&-23&-34&-13&-23&-31&-12&-14/24&12&23&34&13\\
\hline
L22&x&y&z&t&-z&-x&-y&-t&-x&-y&-z&-t&z&x&y&t\\
&23&31&12&14/24&-12&-23&-31&-34&-23&-31&-12&-14/24&12&23&31&34\\
\hline
L23&x&y&z&t&-y&-x&-t&-z&-x&-y&-z&-t&y&x&t&z\\
&23&31&12&14&-31&-23&-24&-12&-23&-31&-12&-14&31&23&24&12\\
\hline
L24&x&y&z&t&-y&-x&-z&-t&-x&-y&-z&-t&y&x&z&t\\
&23&31&12&14&-31&-23&-12&-24&-23&-31&-12&-14&31&23&12&24\\
\end{tabular}
}

Table 2. Directions of the wave front of paths $L21, L22,  L23, L24$ 

\begin{figure*}[htb]
\begin{minipage}{0.47\linewidth}
\begin{center}
\includegraphics[width=4cm,angle=0,clip]{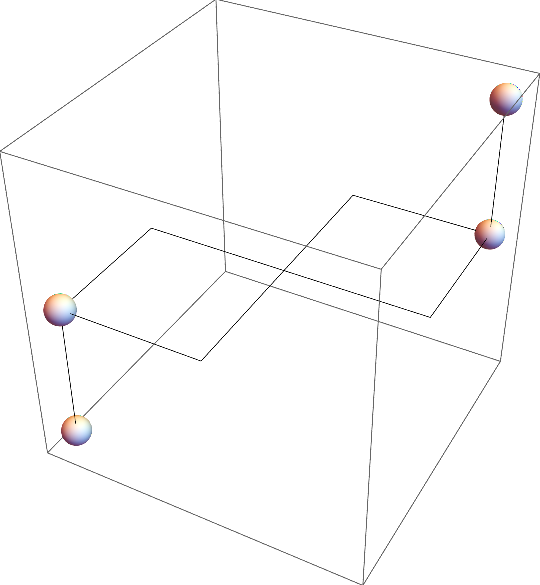}
\end{center}
\end{minipage}
\quad
\begin{minipage}{0.47\linewidth}
\begin{center}
\includegraphics[width=4cm,angle=0,clip]{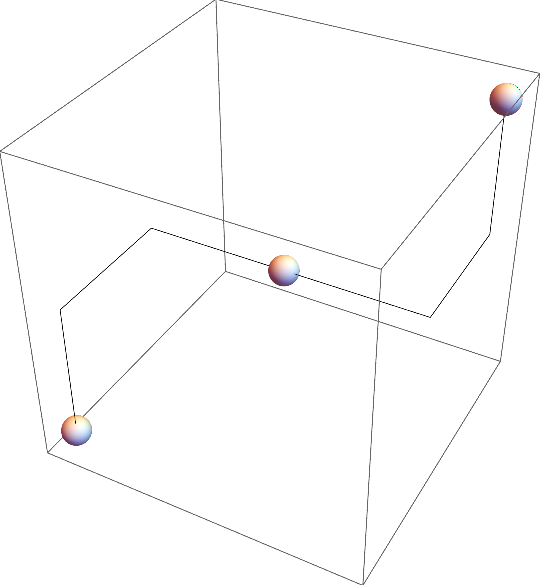}
\end{center}
\end{minipage}
\caption{The path of $L19$(left) and that of $L20$(right). Balls are the places where hysteretic time shift occurs. }\label{l1920}
\end{figure*}

\begin{figure}[htb]
\begin{center}
\includegraphics[width=4cm,angle=0,clip]{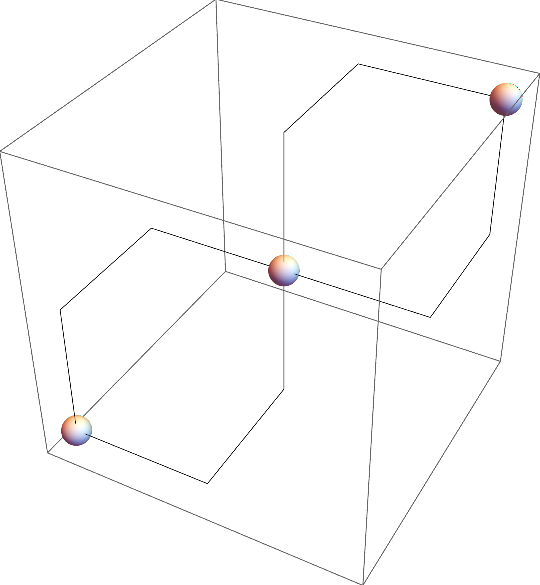}
\end{center}
\caption{The path of $L25$.}\label{l25}
\end{figure}

\begin{figure*}[htb]
\begin{minipage}{0.47\linewidth}
\begin{center}
\includegraphics[width=4cm,angle=0,clip]{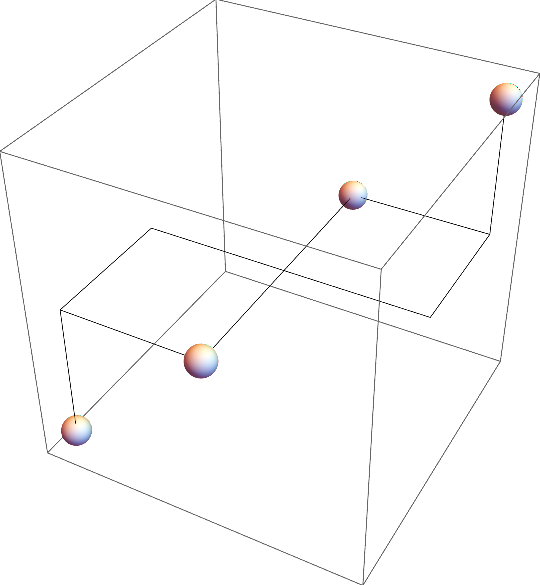}
\end{center}
\end{minipage}
\quad
\begin{minipage}{0.47\linewidth}
\begin{center}
\includegraphics[width=4cm,angle=0,clip]{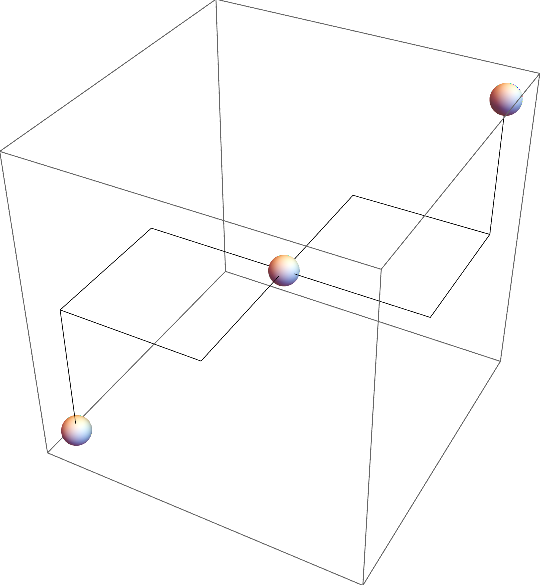}
\end{center}
\end{minipage}
\caption{The path of $L21$(left) and that of  $L22$(right).}\label{l2122}
\end{figure*}

\begin{figure*}[htb]
\begin{minipage}{0.47\linewidth}
\begin{center}
\includegraphics[width=4cm,angle=0,clip]{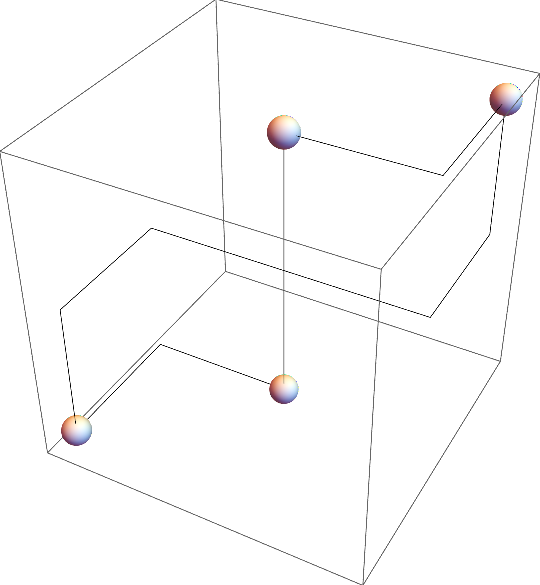}
\end{center}
\end{minipage}
\quad
\begin{minipage}{0.47\linewidth}
\begin{center}
\includegraphics[width=4cm,angle=0,clip]{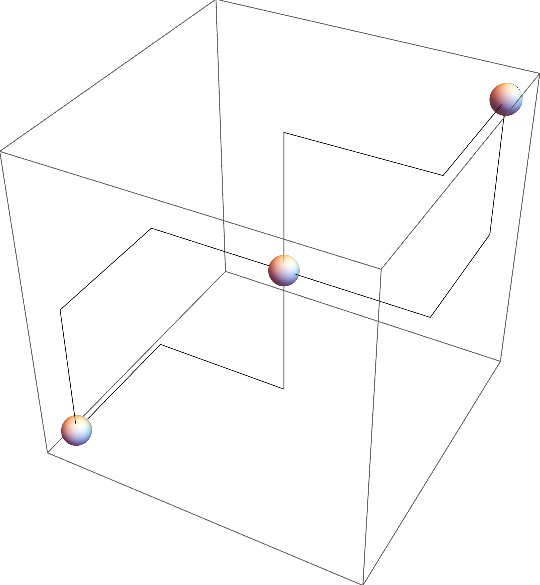}
\end{center}
\end{minipage}
\caption{The path of $L23$(left) and that of $L24$(right).}\label{l2324}
\end{figure*}

The 8 steps of $L19, \cdots, L25$ are shown in \cite{SF23}. The 16 steps of these paths are shown in Figs.\ref{l1920},\ref{l25}, \ref{l2122},\ref{l2324}.  
At balls, time shifts occur. We assume same hysteretic effects occur stochastically in the balls.

The structure of the rest of the article is as follows. In sect.2 we explain the method of obtaining weightfunctions of paths. In sect.3, Markov Chain Monte Carlo is explained. Conclusion and outlook are given in sect.4. 

\section{Optimization of the weight function by Machine Learning techniques}

As in the case of $(2+1)D$, we optimize the weight function of 9 paths ($L19,L20,L21,L21',L22,L22',$ $L23, L24, L25$) in $(3+1)D$ that minimize the path integral action \cite{Feynman48}.

We adopt a cylindrical lattice model, such that 9 paths start from the origin of a space and returns to the origin. The total action becomes 0 when the path returns to the origin. Therefore at steps 7, 8 and 14,15,16 the action of some path becomes 0.

On the input vector $Y=(y_{19}, y_{20},y_{21},y_{21}', y_{22}, y_{22}',y_{23},y_{24},y_{25})^T$, where $^T$ means the transpose, we calculate the integral of action. 
The $L21$ and $L22$ have direction of wave front along $e1e4$ or $e2e4$, and the latter was distinguished by the prime.
\begin{figure}[htb]
\begin{center}
\includegraphics[width=12cm,angle=0,clip]{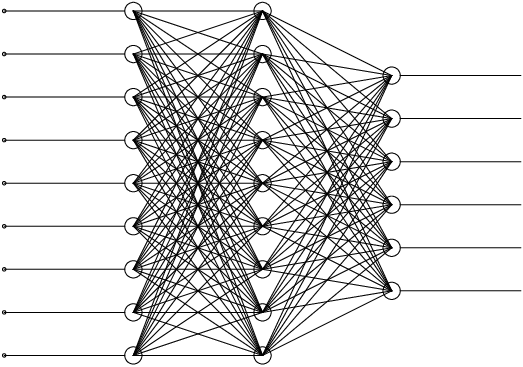}
\end{center}
\caption{ Forward propagating of 9 inputs to 6 outputs through 2 hidden layers in NN. Sources of bias are ignored.}\label{NN}
\end{figure}
The optimal weight function is searched using data of $j_2$ in which action of 9 paths are not zero. For $(u_2/\Delta)/16=j_2=2,3,4,5,6,9,10,11,12,13$, at least one path has non zero action, but action of most paths at $j_2=6$ are zero, and at $j_2=13$, some paths have problem. We choose 6 steps $j_2=4,5,9,10,11,12$ which do not contain 0 action component. In $j_2=9$ the action of the path $L25$ is exceptionally 0, but we include the path for the optimization.
In the case of $j_2=3$, action of all paths are equal and we exclude it for the optimization. 

Following usual ML algorithms, we define the $9\times 9$ matrix $W$ with 51 random numbers between 0 and 1.

In order to calculate transition matrices $X$, we first calculate action integral over $u_1/\Delta=i$ ($1\leq i \leq 255$) at fixed $(u_2/\Delta)/16=j$.

The training vector is prepared by random numbers as $y_1,y_2,\cdots ,y_9$, 18 replicas, and similar validation vector is prepared with 8 repricas.  Repricas mean ordering of random numbers are assigned as in the traveling salesman problem\cite{PM96,KR96}..

We produce 18 random training vectors which consist of 9 elements that specify the weight of $L19,L20,L21,L21',L22,L22',L23, L24$ and $L25$. 
 We make an inner product of the training vector and the 9 component action integral vector at each step. 

Action integrals are calculated as
\begin{eqnarray}
&&x^1_1=\sum_{i=1}^{255}(a19_{4u}(i)+a19_{4d}(i)), \cdots ,x^1_6=\sum_{i=1}^{255}(a19_{12u}(i)+a19_{12d}(i))\nonumber\\
&&x^2_1=\sum_{i=1}^{255}(a20_{4u}(i)+a20_{4d}(i)), \cdots ,x^2_6=\sum_{i=1}^{255} (a20_{12u}(i)+a20_{12d}(i))\nonumber\\
&&x^3_1=\sum_{i=1}^{255}(a21_{4u}(i)+a21_{4d}(i)), \cdots ,x^3_6=\sum_{i=1}^{255}(a21_{12u}(i)+a21_{12d}(i))\nonumber\\
&&x^4_1=\sum_{i=1}^{255}(a21'_{4u}(i)+a21'_{4d})(i),\cdots ,x^4_6=\sum_{i=1}^{255} (a21'_{12u}(i)+a21'_{12d}(i))\nonumber\\
&&x^5_1=\sum_{i=1}^{255}(a22_{4u}(i)+a22_{4d}(i)), \cdots ,x^5_6=\sum_{i=1}^{255}(a22_{12u}(i)+a22_{12d}(i))\nonumber\\
&&x^6_1=\sum_{i=1}^{255}(a22'_{4u}(i)+a22'_{4d}(i)),\cdots ,x^6_6=\sum_{i=1}^{255} (a22'_{12u}(i)+a22'_{12d}(i))\nonumber\\
&&x^7_1=\sum_{i=1}^{255}(a23_{4u}(i)+a19_{4d}(i)), \cdots ,x^7_6=\sum_{i=1}^{255}(a23_{12u}(i)+a23_{12d}(i))\nonumber\\
&&x^8_1=\sum_{i=1}^{255}(a24_{4u}(i)+a20_{4d}(i)), \cdots ,x^8_6=\sum_{i=1}^{255} (a24_{12u}(i)+a24_{12d}(i))\nonumber\\
&&x^9_1=\sum_{i=1}^{255}(a25_{4u}(i)+a21_{4d}(i)), \cdots ,x^9_6=\sum_{i=1}^{255}(a25_{12u}(i)+a25_{12d}(i))
\end{eqnarray}
where $a19_{4u}$ means the action of $L19$ of fixed $u_4/\Delta=16\times 4=64$, that originates from the large eigenvalue, and $a19_{4d}$ means that originates from the small eigenvalue.

The result of the sum $x^1_j y_1+x^2_j y_2+\cdots+x^9_j y_9$ for the 6 $j_2$ show that error bars in the IR region are relatively large, and we optimize the weight function via ML by adopting the affine transformation.
\begin{equation}
X'=X W+b
\end{equation}
where $X$ is the training vector, $b$ is the bias vector  ($b=(b_{19},b_{20},b_{21},b_{21}',b_{22},b_{22}',b_{23},b_{24},b_{25})$), and the matrix $W$ has the form
\begin{equation}
W=\left(\begin{array}{ccccccccc}
\bullet&*&*&*&*&*&0&0&0\\
*&\bullet&*&*&*&*&0&0&0\\
*&*&\bullet&*&*&*&0&0&0\\
*&*&*&\bullet&*&*&0&0&0\\
*&*&*&*&\bullet&*&0&0&0\\
*&*&*&*&*&\bullet&0&0&0\\
0&0&0&0&0&0&\bullet&*&*\\
0&0&0&0&0&0&*&\bullet&*\\
*&*&*&*&*&*&*&*&\bullet\end{array}\right)
\end{equation}

The paths $L19,L20,L25$  returns to the origin in 8 steps. We consider  $L21,L21',L22,L22'$ are correlated with $L19,L20$ and $L25$. The paths $L23, L24$ are correlated with $L25$. In the matrix $W_{ij}$, $1\leq i,j \leq 9$ specify the path. There are 51 * or $\bullet$ where the random number between 0 and 1 will be assigned.

Hysteresis effects induce difference in actions of $L21$ and $L21'$ and $L22$ and $L22'$ at the step 4 and the step 12.
The $L23$ and $L24$ do not contain mixing of $e1e4$ and $e2e4$ and the actions are similar. 

The input vector $Y$ is 9 dimensional and we prepare 18 sets for training and 9 sets for validation. 

From input $Y^{(in)}$ one gets $Z^{(h)}=Y^{(in)} W^{(h)T}+b^{(h)}$ in the hidden layer, and we activate using the sigmoid function as $A^{(h)}=\sigma(Z^{(h)})$, 
\[
\sigma(z)=\frac{1}{1+e^{-z}}
\]
The value $z$ is standardized as \cite{RLM22},
\[
z_{m,n}^{std}=\frac{z_{m,n}-E(z_{m,n})}{Var(z_{m,n})}
\]
 where $E(z_{m,n})$ is the average of the set $z_{m,n}$ and $Var(z_{m,n})$ is the variance  of the set $z_{m,n}$. 

\begin{figure*}[htb]
\begin{minipage}{0.47\linewidth}
\begin{center}
\includegraphics[width=7cm,angle=0,clip]{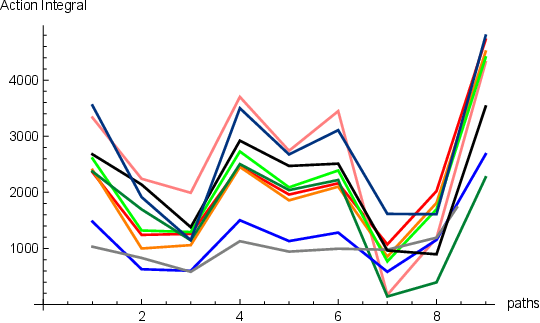}
\end{center}
\end{minipage}
\quad
\begin{minipage}{0.47\linewidth}
\begin{center}
\includegraphics[width=7cm,angle=0,clip]{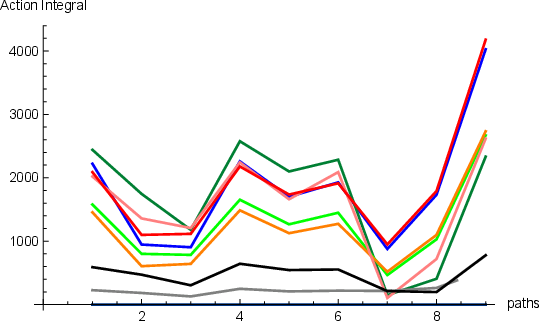}
\end{center}
\end{minipage}
\caption{The weight of output $1_0$(left) and that of output $2_0$(right).  }\label{weight12}
\begin{minipage}{0.47\linewidth}
\begin{center}
\includegraphics[width=7cm,angle=0,clip]{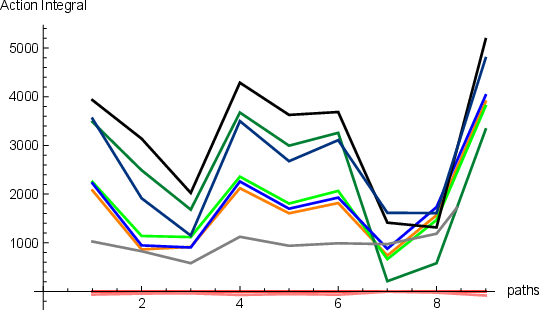}
\end{center}
\end{minipage}
\quad
\begin{minipage}{0.47\linewidth}
\begin{center}
\includegraphics[width=7cm,angle=0,clip]{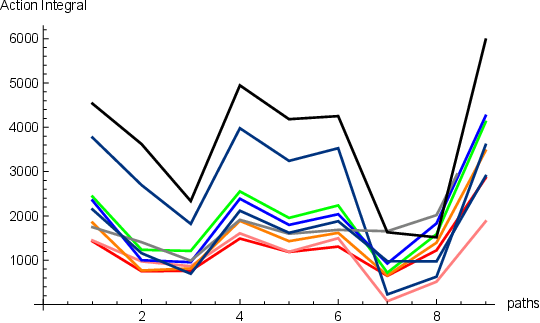}
\end{center}
\end{minipage}
\caption{The weight of output $3_0$(left), and that of output $4_0$(right).  }\label{weight34}
\end{figure*}

\begin{figure*}[htb]
\begin{minipage}{0.47\linewidth}
\begin{center}
\includegraphics[width=7cm,angle=0,clip]{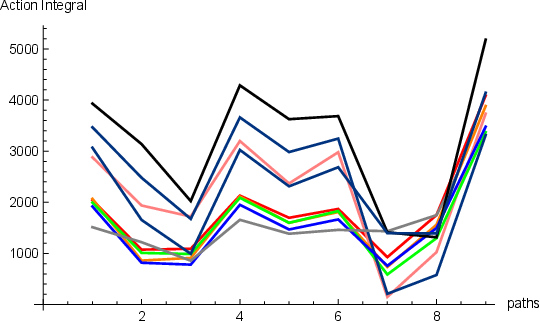}
\end{center}
\end{minipage}
\quad
\begin{minipage}{0.47\linewidth}
\begin{center}
\includegraphics[width=7cm,angle=0,clip]{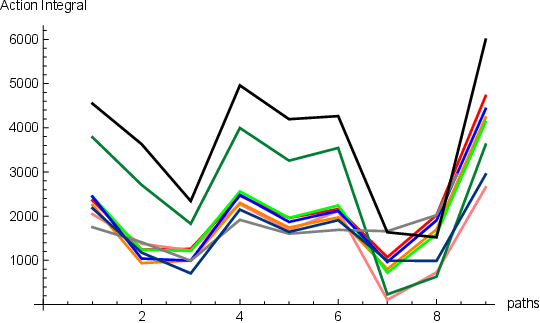}
\end{center}
\end{minipage}
\caption{The weight of output $5_0$(left), and that of $6_0$(right).  }\label{weight56}
\end{figure*}

The amplitude on the output layer is obtained by using $Z^{(x)}=A^{(h)}W^{(out )T}+b^{(x)}$ as $A^{(x)}=\sigma(Z^{(x)})$.
Here $z$ is standardized by 6 output momenta of $(u_2/\Delta)/16=4,5,9,10,11,12$.
In Figs.\ref{weight12},\ref{weight34},\ref{weight56}, the output layer function $A^{(x)}$ of $(u_2/\Delta)/16$ equal 4,5,9,10,11,12, respectively. The color signifies the value of $n$, for a fixed $m$. $(1\leq n\leq 9, 1\leq m\leq 9)$. 
 
Since we want to reduce variations, we take the loss function to be $L=(y-a^{(x)})^2$, where $y$ is the average of $a^{(x)}$ of the one step before.

For three random variables $X$ and $Y$, one defines \cite{Calin20} the conditional entropy $H(Y|X)$ for the probability $p(x_i)=P(X=x_i)$,
$p(y_j)=P(Y=y_j)$, $i=1,\cdots,N$, $j=1,\cdots,M$.
\[
H(Y|X=x_i)=-\sum_{j=1}^M p(y_j|x_i)\log p(y_j|x_i)
\]
and
\begin{eqnarray}
H(Y|X)&=&\sum_{i=1}^N p(x_i)H(Y|X=x_i)\nonumber\\
&=&-\sum_{i=1}^N \sum_{j=1}^M p(x_i)p(y_j|x_i)\log p(y_j|x_i)
\end{eqnarray}
For random variables $X,Y$ and $Z$, the Markov chain $X\to Y\to Z$ satisfies $p(Z|X,Y)=p(Z|Y)$.
It means that the past is essentially conditioned only by the previous variable.

\begin{eqnarray}
\frac{\partial L}{\partial w^{(x)}}&=&2(a^{(x)}-y)\frac{\partial a^{(x)}}{\partial z^{(x)}}\frac{\partial z^{(x)}}{\partial w^{(x)}}\nonumber\\
&=&2(a^{(x)}-y)a^{(x)}(1-a^{(x)}) a^{(h)}
\end{eqnarray}

The bias vector in the output layer is $\eta\frac{\partial L}{\partial b}\sim \eta a^{(x)}(1-a^{(x)}) a^{(h)}$ where $\eta$ is a learning rate.

The weight $w^{(x)}$ is updated via stochastic gradient descent update with a learning rate $\eta$ as
\begin{equation}
w'^{(x)}=w^{(x)}-\eta \frac{\partial L}{\partial w^{(x)}}
\end{equation}


The weight of 9 paths at $j_2=4,5,9,10,11,12$ are shown in Figs.\ref{weight12}, \ref{weight34}, \ref{weight56}.
\begin{figure*}[htb]
\begin{minipage}{0.47\linewidth}
\begin{center}
\includegraphics[width=7cm,angle=0,clip]{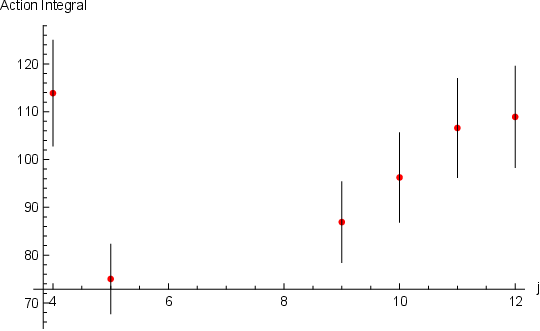}
\end{center}
\end{minipage}
\quad
\begin{minipage}{0.47\linewidth}
\begin{center}
\includegraphics[width=7cm,angle=0,clip]{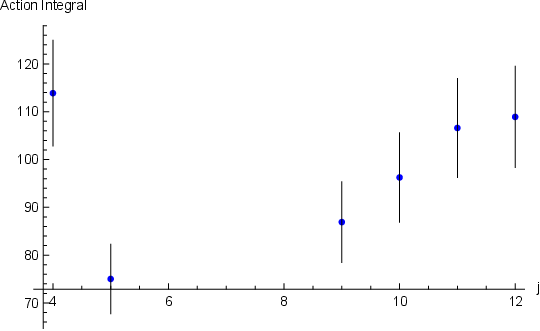}
\end{center}
\end{minipage}
\caption{Action integral of the 1st iteration (left), and that of the 2nd iterations (right). They are identical.
Error-bars originate from 9 random number set.  }\label{actionintegral20}
\end{figure*}

The action integral of $L21'$(4th path) and $L22'$ (6th path) and $L25$ (9th path) are relatively large.

In our model of separating $L19-L22'$and $L23-L25$, the change of action integral by iterations of $A^{(h)}$ and $A^{(x)}$ calculations are not large.  
In the left side of Fig.\ref{actionintegral20}, the integral of the 1st iteration and the 2nd iteration are compared. The latter (blue)  and  the former (red) almost the same.  

In this test run, input is 9 random number sets and output is 6 action integrals. We observed that the output is independent of number of iterations. We need to consider various 9 random number sets and obtain hidden layer sets, choose 51 parameters for the transition between the 9 channel hidden layers and calculate output-layer sets
The optimal weight function could be obtained by searching the set that yields minimal action. 

\section{Markov Chain Monte Carlo}

One defines the transition probability of Markov process $X(t)$ as $F(t,x; u,E)$, which has the following property \cite{ItoK52},

1) For arbitrarily fixed $t,u,x$, it is a measurable distribution of $x\in E$. 

\begin{equation}
F(t,x;t,E)=\left\{\begin{array}{cc}
1&(E\ni x)\\
0&(E\not\ni x)\end{array}\right. 
\end{equation}

2) For a fixed $t,u,E$, $x\in E$ and  $t<u<v$, it satisfies the Chapman equation
\begin{equation}
F(t,x;v,\cdot)\simeq\int F(t,x;u,dy)F(u,y;v,\cdot),\quad (t<u<v)
\end{equation}

One fixes $t,x$ and considers $\varphi_s(E)=F(t,x;s,E)$, $(s\geq t)$, and $f_s(x)=F(t,x;s,E)$, $(t\leq s)$.
We assume that the path in time direction at a fixed lattice point can be approximated by the Markov process, whose transition probability is $F(t,x;s,E)$ and generators $A_t, A^*_t$.

One defines the forward equation
\begin{equation}
\frac{d\varphi_s(E)}{ds}=(A_s^*\varphi_s)(E)
\end{equation}
and the backward equation
\begin{equation}
\frac{df_t}{dt}=-A_t f_t.
\end{equation}

When $X(t)$ is defined at $S=\{1,2,\cdots,r\}$, $A(t)$ is expressed by a matrix $(\alpha_{ij}(t))$ $A^*(t)$ is expressed by its transposed matrix. $\alpha_{ij}(t)$ satisfy stochastic differential equations.

Ito\cite{ItoK52} showed that the stochastic integral equation 
\begin{equation}
X(t)\simeq X(t_0)+\int_{t_0}^t a(\tau, X(\tau))d\tau+\int_{t_0}^t b(\tau,X(\tau))dB(\tau)
\end{equation}
can be applied to Markov processes.
Here $a(\tau, X(\tau))$ and $b(\tau,X(\tau))$ are related to
\begin{equation}
dX(t)=a(t,X(t))dt+b(t,X(t))dB(t),\quad t_0\leq t\leq t_1
\end{equation}
and $B(t)$ is the random walk Wiener process in $t_0\leq t\leq t_1$

$\tau$ originates from the modification of $F(t,x;s,y)\ni f$ to $M(a,b,B,\Xi)\ni f$, $(t,s)\subset (a,b)$ and well definedness of
\begin{equation}
\int_t^\sigma f(\tau,\omega)dB(\tau,\omega),\quad t\leq\sigma\leq s.
\end{equation}.

The stochastic mixing of 9 paths in the hidden layer with hysteresis could be incorporated in $b(t,X(t))dB(t)$, and the path in the 3D space represent the $a(t,X(t))dt$ part. We remark that the Radon-Nykodim's theorem says that in Borel measurable space, additive measurable function $\Phi(E)$ is decomposed to an absolutely continuous function $F(E)$ and singular function $\Psi(E)$\cite{ItoS65}
\[
\Phi(E)=F(E)+\Psi(E).
\] 
In our case proceses containing shifts of Markov time\cite{IM58} contribute to $\Psi(E)$.

 In the Kolmogorov's approach\cite{ItoK52,Feller57}, variance $V(X)$ and mean $E(X)$ are defined in Borel measure space
\begin{eqnarray}
V(X)&=&\int_\Omega (X(\omega)-E(X))^2 dP(\omega),\nonumber\\
E(X)&=&\sum_{n=1}^4 a_n P(X=a_n)
\end{eqnarray}
in the probability space $\Omega({\bf A},P)$. The average $E(X)$ and the variance $V(X)$ are defined by the Lebesgue integral\cite{ItoS65}.

Symmetric random walks of a particle in 1D and 2D, and 3D have large qualitative difference. The propability of a particle returns to the original position in 1D and 2D is 1, but that in 3D is about 0.35\cite{Feller57}. 

Computation of path integral of QCD using Markov Chain Monte Carlo method and Machine Learning techniques are reported in \cite{AABHKRLSU23}.

\section{Data transfer between hidden layers}
In this section, we include data transfer between two hidden layers through the coincidence of time shift positions of paths. The technique  is similar to the treatment of Recurrent Neural Network (RNN)\cite{RLM22}. 

We orepare 18 input layer 9D vector at $t=0$: ${\bf x}^{0}$ and hidden layer vector ${\bf h}^{(0)}$.
In RNN, the weight matrix is taken as
\begin{itemize}
\item ${\bf W}_{xh}$: The weight matrix between the input ${\bf x}^{(i)}$  and the hidden layer ${\bf h}$.
\item ${\bf W}_{hh}$: The weight matrix associated with the recurrent edge.
\item ${\bf W}_{ho}$: The weight matrix between the hidden layer and output layer.
\end{itemize}
In \cite{RLM22}, the concatenated weight matrix ${\bf W}_h=[{\bf W}_{xh}:{\bf W}_{hh}]$, 
\begin{equation}
{\bf W}_h=\left(\begin{array}{ccccccccc ccccccccc}
\bullet&*&*&*&*&*&0&0&0 &\circ&\circ&\circ&\circ&\circ&\circ&\circ&\circ&\circ\\
*&\bullet&*&*&*&*&0&0&0 &\circ&\circ&\circ&\circ&\circ&\circ&\circ&\circ&\circ\\
*&*&\bullet&*&*&*&0&0&0 &\circ&\circ&\circ&\circ&\circ&\circ&\circ&\circ&\circ\\
*&*&*&\bullet&*&*&0&0&0 &\circ&\circ&\circ&\circ&\circ&\circ&\circ&\circ&\circ\\
*&*&*&*&\bullet&*&0&0&0 &\circ&\circ&\circ&\circ&\circ&\circ&\circ&\circ&\circ\\
*&*&*&*&*&\bullet&0&0&0 &\circ&\circ&\circ&\circ&\circ&\circ&\circ&\circ&\circ\\
0&0&0&0&0&0&\bullet&*&* &\circ&\circ&\circ&\circ&\circ&\circ&\circ&\circ&\circ\\
0&0&0&0&0&0&*&\bullet&* &\circ&\circ&\circ&\circ&\circ&\circ&\circ&\circ&\circ\\
*&*&*&*&*&*&*&*&\bullet &\circ&\circ&\circ&\circ&\circ&\circ&\circ&\circ&\circ\end{array}\right),
\end{equation}
the state vector
\[
{\bf x}=(x_1^{(t)},x_2^{(t)},\cdots,x_7^{(t)},x_8^{(t)},x_9^{(t)},h_1^{(t-1)},h_2^{(t-1)},\cdots,h_7^{(t-1)},h_8^{(t-1)},h_9^{(t-1)})^T,
\]
and the activation of the output unit 
\begin{equation}
{\bf o}^{(t)}=\sigma({\bf W}_{ho}({\bf h}^{(t)})).
\end{equation}
are used. 

However, since evaluation of ${\bf b}_h$ is not trivial, we calculate
\begin{equation}
{\bf h}^{(t)}=\sigma([{\bf W}_{xh}:{\bf W}_{hh}]\left[\begin{array}{c}
{\bf x}^{(t)}+{\bf b}_x\\
{\bf h}^{(t-1)}\end{array}\right] ).
\end{equation}

The bias vectors is
\[
 {\bf b}_x=(b_x^1,b_x^2,\cdots,b_x^7,b_x^8,b_x^9)^T,
\]

We take the activations of the hidden units at the time $t=5,8,9,13,16,17$ as
\begin{equation}
{\bf h}^{(t)}=\sigma({\bf z}^{(t)})=\sigma({\bf W}_{xh}( {\bf x}^{(t)}+{\bf b}_x)+{\bf W}_{hh}({\bf h}^{(t-1)})).
\end{equation}

Since we have input vectors ${\bf x}^{(t)}$ and ${\bf h}^{(t-1)}$, we don't concatenate matrices, and calculate ${\bf b}_x$.

The weight matrix of hidden layers ${\bf W}_{hh}^{(t)}$ depends on time $(t)$. We assume that the element of ${\bf W}_{hh}^{(t)}$, denoted as $W_{m,n}^{(t)}$ 
is not 0 when the direction of the path $m$: $e_i e_4$ and that of $n$ : $e_j e_4$ coincide. At $t=4,7,8,12,15$ and 16, there are time shift points between different paths.  

The interactions between hidden layers depends on whether time shift occurs or not.    
\begin{equation} 
{\bf W}_{hh}^{(4)}=\left(\begin{array}{ccccccccc}
*&*&0&*&0&*&0&0&*\\
*&*&0&*&0&*&0&0&*\\
0&0&*&0&*&0&*&*&0\\
*&*&0&*&0&*&0&0&*\\
0&0&*&0&*&0&*&*&0\\
*&*&0&*&0&*&0&0&*\\
0&0&*&0&*&0&*&*&0\\
0&0&*&0&*&0&*&*&0\\
*&*&0&*&0&*&0&0&*\end{array}\right)\quad {\bf W}_{hh}^{(7)}=\left(\begin{array}{ccccccccc}
0&0&0&0&0&0&0&0&0\\
0&0&0&0&0&0&0&0&0\\
0&0&*&0&0&0&0&0&0\\
0&0&0&*&0&0&0&0&0\\
0&0&0&0&0&0&0&0&0\\
0&0&0&0&0&0&0&0&0\\
0&0&0&0&0&0&*&0&*\\
0&0&0&0&0&0&0&0&0\\
0&0&0&0&0&0&*&0&*\end{array}\right)
\end{equation}

\begin{equation} 
{\bf W}_{hh}^{(8)}=\left(\begin{array}{ccccccccc}
0&0&0&0&0&0&0&0&0\\
0&*&0&0&0&0&0&*&*\\
0&0&0&0&0&0&0&0&0\\
0&0&0&0&0&0&0&0&0\\
0&0&0&0&*&0&0&0&0\\
0&0&0&0&0&*&0&0&0\\
0&0&0&0&0&0&*&0&0\\
0&*&0&0&0&0&0&*&*\\
0&*&0&0&0&0&0&*&*\end{array}\right)\quad {\bf W}_{hh}^{(12)}=\left(\begin{array}{ccccccccc}
*&*&0&*&0&*&0&0&*\\
*&*&0&*&0&*&0&0&*\\
0&0&*&0&*&0&*&*&0\\
*&*&0&*&0&*&0&0&*\\
*&*&0&*&0&*&0&0&0\\
*&*&0&0&0&*&0&0&*\\
0&0&*&0&*&0&*&*&0\\
0&0&*&0&*&0&*&*&0\\
*&*&0&*&0&*&0&0&*\end{array}\right)
\end{equation}

\begin{equation} 
{\bf W}_{hh}^{(15)}=\left(\begin{array}{ccccccccc}
0&0&0&0&0&0&0&0&0\\
0&0&0&0&0&0&0&0&0\\
0&0&*&0&0&0&0&0&0\\
0&0&0&*&0&0&*&0&0\\
0&0&0&0&0&0&0&0&0\\
0&0&0&0&0&0&0&0&0\\
0&0&0&*&0&0&0&0&0\\
0&0&0&0&0&0&0&0&0\\
0&0&0&0&0&0&0&0&0\end{array}\right)\quad {\bf W}_{hh}^{(16)}=\left(\begin{array}{ccccccccc}
0&0&0&0&0&0&0&0&0\\
0&*&0&0&0&0&0&*&*\\
0&0&0&0&0&0&0&0&0\\
0&0&0&0&0&0&0&0&0\\
0&0&0&0&*&*&0&0&0\\
0&0&0&0&*&*&0&0&0\\
0&0&0&0&0&0&0&0&0\\
0&*&0&0&0&0&0&*&*\\
0&*&0&0&0&0&0&*&*\end{array}\right)
\end{equation}
The low and the column are in the order $L19,L20,L21,L21',L22,L22',L23,L24,L25$ and $*$ indicate time shifts in the own path occur or time shifts that cause mixing of paths occur. The path $L21$ and $L21'$ do not mix, similarly $L22$ and $L22'$ do not mix. We are trying to optimize the path using 18 training sets and 8 validation sets.

\subsection{Elman Recurrent Neural Network method}
When there are two vectors ${\bf x}^{(t)}$ and ${\bf h}^{(t-1)}$ as the input, the optimal bias vector ${\bf b}_h$ is hard to define. 
In the Elman Recurrent Neural Network (ERNN) \cite{Bianchi17},  the bias vector for ${\bf x}$ denoted as ${\bf b}_x$, and that for ${\bf h}$ denoted as ${\bf b}_h$ are calculated using $\frac{\partial L}{\partial {\bf W}_{xh}}$ and using $\frac{\partial L}{\partial {\bf W}_{hh}}$, respectively. In ERNN, there is no back propagation, adopted in the NN of \cite{RLM22}.

Choosing the expected value $a_k^{(x)}, a_k^{(h)}$ as the means of the 18 samples of ${\bf W}^{(x)}{\bf x}$ and ${\bf W}^{(h)}{\bf h}$, respectively, 
the loss function is 
\begin{equation}
L=\sum_k (y_x-a_k^{(x)})^2 +\sum_k(y_h-a_k^{(h)})^2
\end{equation}
\begin{eqnarray}
&&\frac{\partial L}{\partial a_k^{(x)}}=-2(a_k^{(x)}-y_x), \quad \frac{\partial L}{\partial a_k^{(h)}}=-2(a_k^{(h)}-y_h) \nonumber\\
&&\frac{\partial a_k^{(x)}}{\partial W_{k,k}^{(x)}}=\frac{\partial}{\partial W_{k,k}^{(x)}}(a_k^{(h)}W_{k,k}^{(x)}+b_k^{(x)})=a_k^{(h)},\quad
\frac{\partial a_k^{(h)}}{\partial W_{k,k}^{(h)}}=0\nonumber\\
\end{eqnarray}
We ignore variation of ${\bf W}_{hh}^{(t)}$ and choose ${\bf b}_h^{(t)}=0$.
For the activation we adopt the logistic sigmoid function. The weight function ${\bf W}^{(x)}$ is modified as
\begin{equation}
W_{k,k}^{(x)}:=W_{k,k}^{(x)}-\eta\langle\frac{\partial L}{\partial W_{k,k}^{(x)}}\rangle,
\end{equation}
where $\frac{\partial L}{\partial W_{k,k}^{(x)}}=-2(a_k^{(x)}-y_x) a_k^{(h)}$, 
 and $\langle \frac{\partial L}{\partial W} \rangle$ is an average over 18 samples of  $\frac{\partial L}{\partial W}$ in the training process and 8 samples in the validation process.  The learning rate parameter $\eta$ is chosen to be 0.01.

With the new ${\bf W}^{(x)}$ we calculate the new ${\bf z}_h$ and ${\bf h}$
\begin{equation}
{\bf z}_h^{(t)}={\bf W}^{(x)}( {\bf x}^{(t)}+{\bf b}_x^{(t)})+{\bf W}^{(h)}({\bf h}^{(t-1)}),\quad {\bf h}^{(t)}=\sigma({\bf z}_h^{(t)})
\end{equation}
or when ${\bf W}^{(h)}$ is a null matrix
\begin{equation}
{\bf z}_h^{(t)}={\bf W}^{(x)}( {\bf x}^{(t)}+{\bf b}_x^{(t)}),
\end{equation}
and calculate the output
\begin{equation}
{\bf z}_{out}^{(t)}={\bf W}_{ho}({\bf z}_h^{(t)}), \quad {\bf y}_{out}^{(t)}=\sigma({\bf z}_{out}^{(t)}).
\end{equation}

We start from step 4, ${\bf h}^{(4)}=\sigma({\bf W}_{xh}({\bf x}^{(4)}+{\bf b}_x^{(4)}))$, and randomly produced 9D vector ${\bf h}^{(a)}$. 
We calculate 
\begin{equation}
{\bf W}_{xh}{\bf b}_x^{(4)}=(2(x_1-m_1)h^{(a)}_1,\cdots, 2(x_9-m_9)h^{(a)}_9)^T,
\end{equation}
where $m_k$ is the $k$th component of the mean of ${\bf W}_{xh} {\bf x}^{(4)}$. $h^{(a)}_k$ is the $k$th component of the hidden layer vector ${\bf h}^{(a)}$.

At the step 5, ${\bf h}^{(5)}=\sigma({\bf W}_{xh}({\bf x}^{(5)}+{\bf b}_x^{(5)})+{\bf W}_{hh}^{(4)}({\bf h}^{(a)}))$, where 
\begin{equation}
{\bf W}_{xh}{\bf b}_x^{(5)}=(2( x_1-m_1)h^{(a)}_1,\cdots,2 (x_9-m_9)h^{(a)}_9)^T,\nonumber\\
\end{equation}
and $\sigma({\bf W}_{xh}{\bf b}_x)$ is calculated for evaluating the necessary shifts.

At the step 6, ${\bf h}^{(6)}=\sigma({\bf W}_{xh}({\bf x}^{(6)}+{\bf b}_x^{(6)}))$. 

At the step 7, ${\bf h}^{(7)}=\sigma({\bf W}_{xh}({\bf x}^{(7)}+{\bf b}_x^{(7)}))$. 

At the step 8, ${\bf h}^{(8)}=\sigma({\bf W}_{xh}({\bf x}^{(8)}+{\bf b}_x^{(8)})+{\bf W}_{hh}^{(7)}({\bf h}^{(b)})), \quad{\bf h}^{(b)}=\sigma({\bf W}_{hh}^{(7)} {\bf h}^{(a)}).$

At the step 9, ${\bf h}^{(9)}=\sigma({\bf W}_{xh}({\bf x}^{(9)}+{\bf b}_x^{(9)})+{\bf W}_{hh}^{(8)}({\bf h}^{(c)})), \quad{\bf h}^{(c)}=\sigma({\bf W}_{hh}^{(8)} {\bf h}^{(b)}).$

At the step 10, ${\bf h}^{(10)}=\sigma({\bf W}_{xh}({\bf x}^{(10)}+{\bf b}_x^{(10)}))$. 

At the step 11, ${\bf h}^{(11)}=\sigma({\bf W}_{xh}({\bf x}^{(11)}+{\bf b}_x^{(11)}))$. 

At the step 12, ${\bf h}^{(12)}=\sigma({\bf W}_{xh}({\bf x}^{(12)}+{\bf b}_x^{(12)}))$. 

At the step 13, ${\bf h}^{(13)}=\sigma({\bf W}_{xh}({\bf x}^{(13)}+{\bf b}_x^{(13)})+{\bf W}_{hh}^{(12)}({\bf h}^{(d)})), \quad{\bf h}^{(d)}=\sigma({\bf W}_{hh}^{(12)} {\bf h}^{(c)}).$

At the step 14  ${\bf h}^{(14)}=\sigma({\bf W}_{xh}({\bf x}^{(14)}+{\bf b}_x^{(14)}))$. 

At the step 15 ${\bf h}^{(15)}=\sigma({\bf W}_{xh}({\bf x}^{(15)}+{\bf b}_x^{(15)}))$. 

At the step 16, ${\bf h}^{(16)}=\sigma({\bf W}_{xh}({\bf x}^{(16)}+{\bf b}_x^{(16)})+{\bf W}_{hh}^{(15)}({\bf h}^{(e)})), \quad{\bf h}^{(e)}=\sigma({\bf W}_{hh}^{(15)} {\bf h}^{(d)}).$

At the step 17,  ${\bf h}^{(17)}=\sigma({\bf W}_{xh}({\bf x}^{(17)}+{\bf b}_x^{(17)})+{\bf W}_{hh}^{(16)}({\bf h}^{(f)})),\quad{\bf h}^{(f)}=\sigma({\bf W}_{hh}^{(16)} {\bf h}^{(e)}).$

At the step 18  ${\bf h}^{(18)}=\sigma({\bf W}_{xh}({\bf x}^{(18)}+{\bf b}_x^{(18)}))$.

At the step 19 ${\bf h}^{(19)}=\sigma({\bf W}_{xh}({\bf x}^{(19)}+{\bf b}_x^{(19)}))$.

At the step 20, and at any time $t$, we calculate the output of 6 selected epochs
\begin{equation}
{\bf y}^{(t)}={\bf W}_{ho}{\bf h}^{(t)}.
\end{equation}
where
\begin{equation}
{\bf W}_{ho}=\left(\begin{array}{ccccccccc}
*&*&*&*&*&*&*&*&*\\
*&*&*&*&*&*&*&*&*\\
*&*&*&*&*&*&*&*&*\\
*&*&*&*&*&*&*&*&*\\
*&*&*&*&*&*&*&*&*\\
*&*&*&*&*&*&*&*&*\end{array}\right)
\end{equation}
is calculated from the action of 9 paths, and
\begin{equation}
{\bf h}^{(t)}=\sigma({\bf W}_{xh}({\bf x}^{(t)}+{\bf b}_x^{(t)})+{\bf W}_{hh}^{(t-1)}({\bf h}^{(t-1)})).
\end{equation}

The iterations of the step 4 to the step 20 continue until the outputs of 18 training samples and 8 validation samples become close together.


Hysteresis increases the action. Parameters of ${\bf W}_{hh}^{(t)}$ may be adjusted if experimental data are available. In the present work, 
 non-zero random number denoted by $*$ are created by Mathematica module $RandomReal[1,k]$, where $k$ is the number of $*$ in the matrix ${\bf W}_{hh}^{(t)}$. 
\begin{figure}[htb]
\begin{center}
\includegraphics[width=12cm,angle=0,clip]{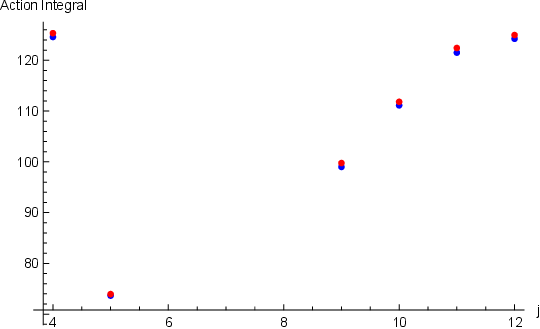}
\end{center}
\caption{Action integral of the average of 18 training samples with fixed ${\bf W}_{hh}$ are blue points and varying ${\bf W}_{hh}$ at each round are red points. The coordinate $"j"$ represents the step $j+16k$ $(k=0,\cdots, {\rm itemax}-1)$. The ordinate "Action integral" is the average over itemax data..}\label{RNN8-20}
\end{figure}

We performed a test run of 20 and 200 iterations of 16 steps, that is starting from the step 4 to the step 16, and return back to the step 1 and proceed until the step 3 makes one round. The Fig.\ref{RNN8-20} is the result of erforming 20 iteration using 18 training samples and 8 validation samples. 
The difference of Action Integral of training samples and validation samples was negligible.

We estimated hysteresis effects by comparing Action Integral using a fixed ${\bf W}_{hh}$ through all rounds and creating ${\bf W}_{hh}$ at each round.  Fig.\ref{RNN8-20} shows the Action Integral of the latter (red points) are slightly larger than the former (blue points). 

\subsection{Long-Short Term Memory layer method}
In recent RNN, long-short term memory (LSTM) layer are used in order to avoid vanishing gradient problem\cite{RLM22,GSC00}.
Although we do not adopt the method in the present work, we explain the method.

One introduces the forget gate ${\bf f}_t$, inputgate ${\bf i}_t$ and thecandidate value $\tilde{\bf C}_t$ 
\begin{eqnarray}
&&{\bf f}_t=\sigma({\bf W}_{xf}{\bf x}^{(t)}+{\bf W}_{hf}^{(t)}{\bf h}^{(t-1)}+{\bf b}_f)\nonumber\\
&&{\bf i}_t=\sigma({\bf W}_{xi}{\bf x}^{(t)}+{\bf W}_{hi}^{(t)}{\bf h}^{(t-1)}+{\bf b}_i )\nonumber\\
&&\tilde{\bf C}_t=\tanh({\bf W}_{xc}{\bf x}^{(t)}+{\bf W}_{hc}^{(t)}{\bf h}^{(t-1)}+{\bf b}_c)
\end{eqnarray}

In our case, ${\bf W}_{xf}={\bf W}_{xi}={\bf W}_{xc}, {\bf W}_{hf}^{(t)}={\bf W}_{hi}^{(t)}={\bf W}_{hc}^{(t)}$ are random matrices, ${\bf x}^{(t)}$ and ${\bf h}^{(t)}$ are 9D vectors. 

The cell state at time $t$ is defined as
\begin{equation}
{\bf C}^{(t)}=({\bf C}^{(t-1)}\odot {\bf f}_t)\oplus({\bf i}_t\odot \tilde{\bf C}_t),
\end{equation}
and the output gate at $t=5,8,9,13,16,17$ are
\begin{equation}
{\bf o}^{(t)}=\sigma({\bf W}_{xo}{\bf x}^{(t)}+{\bf W}_{ho}^{(t)}{\bf h}^{(t-1)}+{\bf b}_o).
\end{equation}
The outputs at other epochs  are
\begin{equation}
{\bf o}^{(t)}=\sigma({\bf W}_{xo}{\bf x}^{(t)}+{\bf b}_o).
\end{equation}

The hidden unit at $t$ is
\begin{equation}
{\bf h}^{(t)}={\bf o}^{(t)}\odot \tanh({\bf C}^{(t)}).
\end{equation}
$\odot$ denotes the element wise multiplication which is known as Hudmard product\cite{HBB20}.
The loss function of ${\bf x}^{t}$ is the difference between the input ${\bf h}^{(t)}$ and the output $\hat{\bf o}^{(t)}\cdot \tanh({\bf C}^{(t)})$.

The loss function of ${\bf h}^{(t)}$ is evaluated by using softmax function.
\begin{equation}
S_i=\frac{e^{y_i}}{\sum_k e^{y_k}}, \quad \ell=-\log S_m
\end{equation} 
The derivative of $S_i$ with respect to $y_j$ is 
\begin{equation}
\frac{\partial S_i}{\partial y_j}=\left\{\begin{array}{cc}
S_i(1-S_j)&j=i\\
-S_iS_j &j\ne i \end{array}\right.
\end{equation}

Derivative of $\tanh(s)$ with respect to $s$ is 
\begin{equation}
\frac{\partial \tanh(s)}{\partial s}=1-\tanh^2(s).
\end{equation}
The iteration stop condition is $({\bf W}_{ho}{\bf h}^{(t)}-{\bf W}_{xo}{\bf x}^{(t)})^2<\epsilon$, where $\epsilon$ is a positive small real number.

The bias vectors ${\bf b}_f, {\bf b}_i$ are chosen to be 0, ${\bf b}_c$ is $\eta( a^{(x)}(1-a^{(x)})a^{(h)}+ a^{(h)}(1-a^{(h)})a^{(x)})<\epsilon$.
The application of LSTM, its bi-directional extension (Bi-LSTM) and gated recurrent neural network (GRU)\cite{CGCB14,Bianchi17} are left for the future.

\section{Conclusion and outlook}
We showed that the weight function of paths defined by the fixed point action can be optimized by the Elman RNN method. The optimal weight functioon of the C-type fixed point actions which contain hysteresis effect can be simulated by using the biquaternion basis. We observed stablility of the action on the output layer produced from the hidden layers. Thee are other RNN methods which are left for the future study.

 As an extension of the (3+1)D model, mappings in the (4+1)D : $j({\mathcal A}_{4,1})\sim M_2({\bf H})\oplus M_2({\bf H})$ can be considered by bi-quaternion matrices.
\[
j({\mathcal A}_{4,1})=\left(\begin{array}{cc}
x_2{\bf i}+x_3{\bf j}+x_4{\bf k}&-x_1+x_5\\
x_1+x_5&-x_2{\bf i}-x_3{\bf j}-x_4{\bf k}\end{array}\right).
\]
The determinant of $j({\mathcal A}_{4,1})$ is $x_2^2+x_3^2+x_4^2+x_5^2-x_1^2$.  When we identify it as $-X_0^2$ and transform $x_2\to X_1$, $x_3\to X_2$, $x_4\to X_3$, $x_1\to X_4$ and $x_5\to X_5$, we obtain the relation $X_1^2+X_2^2+X_3^2+X_5^2-X_4^2=-X_0^2$, or $X_0^2+X_1^2+X_2^2+X_3^2=X_4^2-X_5^2$, that Dirac derived in the Lorentz transformation of quaternions\cite{Dirac45}.

In the light front quantization of QCD by Srivastava and Brodsky\cite{SB01},  a fixed light-front time $\tau=(t-z/c)/\sqrt{2}$ is introduced.
The light-front time corresponds to $(X_4-X_5)/\sqrt 2$. For massless particle, propagators are doubly transverse, i.e. with respect to the gauge direction $n_\mu$ and the chilarity direction $k_\mu$.

 The two $M_2({\bf H})$ represent TR symmetric physical fields, and the BRST ghost fields\cite{BRS75} are decoupled.

L\"uscher\cite{Luescher99} discussed Abelian chiral gauge theories on the lattice using Dirac spinors which consist of two Weyl spinors.
Fermion expectation values of any product $\mathcal O$ of fields are obtained as
\begin{equation}
\langle{\mathcal O}\rangle_F=w[m]\int D[\psi]D[\bar\psi]{\mathcal O}e^{-S_F},
\end{equation}
where the fermion action $S_F$ is asuumed to take the form
\begin{equation}
S_F=\sum_{k,j}\bar c_k M_{kj} c_j, \quad M_{kj}=\sum_{x\in \Gamma}\bar v_k(x)D v_j(x).
\end{equation}
The integration measures are $D[\bar \psi]=\Pi_k d\bar c_k,\quad \bar\psi(x)=\sum_k \bar c_k \bar v_k(x)$.

The weight function $w[m]$ is complex and depends on the presence of zero modes. There is an argument of considering a domain wall in $(4+1)D$ space-time \cite{Kaplan24,KS24}.

 In Klebanov's gauge theory \cite{KT00,Klebanov23}, and Chemtob's theory \cite{Chemtob22}, $S^5\sim S^2\times S^3\sim T^{1,1}\sim S^3_1\times S^3_2/U(1)$ i.e. product of two quaternions modulus $U(1)_R$ symmetry. Quaternion Field Theory proposed by Adler \cite{Adler85} and his extension in the frame work of gauge theory \cite{Adler94} has new progress.

 Quaternion and bi-quaternion basis model can be used not only for NDT, but also for QCD lattice simulations. For getting the optimal solution, ML techniques can be applied. Nonlinearity and hysteresis could be explored in these basis.

\leftline{\bf Acknowledgments}

S.F. thanks the Laboratory for Industrial Research (Nissanken) for the financial aid to the travel expense to INSA Centre Val de Loire in November 2024 and  Prof. S.J. Brodsky for helpful communication. 
The numerical calculation was done using Mathematica of the Wolfram Research installed on a workstation of the faculty of science and engineering of Teikyo University.  S.F. is grateful to Prof. M. Arai and Prof. K. Hamada for the permission.

\vskip 0.5 true cm


\end{document}